\documentclass[sigconf, nonacm, pdfa]{acmart}

\AtBeginDocument{%
  }

\setcopyright{acmcopyright}
\copyrightyear{2024}
\acmYear{2024}
\acmDOI{XXXXXXX.XXXXXXX}

\acmConference[Seattle '23]{Seattle '23: ACM SIGMOD/PODS International Conference on Management of Data}{June 13--23, 2023}{Seattle, USA}
\acmPrice{15.00}
\acmISBN{XXX-X-XXXX-XXXX-X/XX/XX}
\graphicspath{{images/}}
\usepackage{array}
\usepackage{amsmath}
\usepackage{algorithmic}
\usepackage[ruled,vlined]{algorithm2e}
\usepackage{pgfplots,pgfplotstable}  
\usepackage{booktabs}
\usepackage{balance}
\usepackage{multirow}
\usepackage{subcaption}
\usepackage{empheq}
\usepackage{mathtools, nccmath}
\usepackage[a-2b]{pdfx}

\pgfplotsset{compat=newest}

\usepackage{xcolor}
\usepackage{multirow}  
\usepackage{adjustbox} 
\usepackage{optidef}   

\newcommand{\mynote}[1]{}

\newcommand{\yz}[1]{\textcolor{blue}{\mynote{YZ:~#1}}}
\newcommand{\al}[1]{\textcolor{blue}{\mynote{AL:~#1}}}
\newcommand{\edit}[1]{{\color{black}#1}}

\DeclarePairedDelimiter{\nint}\lfloor\rceil
\newcommand{\mypar}[1]{\smallskip\smallskip\noindent\textbf{#1.}\xspace}
\newcommand{\sysname}{Intelligent Pooling\xspace}
\newcommand{\companyname}{Microsoft\xspace}
\newcommand{\msft}{Microsoft\xspace}

\newcommand{\fabric}{Fabric\xspace}

\newcommand\vldbdoi{10.14778/3654621.3654629}
\newcommand\vldbpages{1618 - 1627}

\newcommand\vldbvolume{17}
\newcommand\vldbissue{7}
\newcommand\vldbyear{2024}

\newcommand\vldbauthors{\authors}
\newcommand\vldbtitle{\shorttitle} 

\newcommand\vldbpagestyle{empty} 

\begin{document}

\title{\sysname: Proactive Resource Provisioning in Large-scale Cloud Service}


\author{Deepak Ravikumar }
\authornote{Authors contributed equally to this research.}
\authornote{Work done while at Microsoft}
\affiliation{%
 \institution{Purdue University, USA}
}
\email{dravikum@purdue.edu}

\author{Alex Yeo}
\authornotemark[1]
\authornotemark[2]
\affiliation{%
 \institution{Netflix, USA}
}
\email{alexsyeo@gmail.com}

\author{Yiwen Zhu}
\authornotemark[1]
\affiliation{%
 \institution{Microsoft, USA}
}
\email{yiwzh@microsoft.com}

\author{Aditya Lakra}
\affiliation{%
 \institution{Microsoft, USA}
}
\email{adityalakra@microsoft.com}

\author{Harsha Nagulapalli}
\affiliation{%
 \institution{Microsoft, USA}
}
\email{hanagula@microsoft.com}

\author{Santhosh Ravindran}
\affiliation{%
 \institution{Microsoft, USA}
}
\email{saravi@microsoft.com}

\author{Steve Suh}
\affiliation{%
 \institution{Microsoft, USA}
}
\email{stsuh@microsoft.com}

\author{Niharika Dutta}
\affiliation{%
 \institution{Microsoft, USA}
}
\email{nidutta@microsoft.com}

\author{Andrew Fogarty}
\affiliation{%
 \institution{Microsoft, USA}
}
\email{anfog@microsoft.com}

\author{Yoonjae Park}
\affiliation{%
 \institution{Microsoft, USA}
}
\email{yoonjae.park@microsoft.com}

\author{Sumeet Khushalani}
\affiliation{%
 \institution{Microsoft, USA}
}
\email{sukhusha@microsoft.com}

\author{Arijit Tarafdar} 
\affiliation{%
 \institution{Microsoft, USA}
}
\email{arijitt@microsoft.com}

\author{Kunal Parekh} 
\affiliation{%
 \institution{Microsoft, India}
}
\email{kunalparekh@microsoft.com}

\author{Subru Krishnan}
\affiliation{%
 \institution{Microsoft, Spain}
}
\email{subru@microsoft.com}





\renewcommand{\shortauthors}{Ravikumar, Yeo and Zhu, et al.}

\begin{abstract}
The proliferation of big data and analytic workloads has driven the need for cloud compute and cluster-based job processing. With Apache Spark, users can process terabytes of data at ease with hundreds of parallel executors. 
Providing low latency access to Spark clusters and sessions is a challenging problem due to the large overheads of cluster creation and session startup. In this paper, we introduce \sysname, a system for proactively provisioning compute resources to combat the aforementioned overheads. 
Our system (1) predicts usage patterns using an innovative hybrid Machine Learning (ML) model with low latency and high accuracy; and (2) optimizes the pool size dynamically to meet customer demand while reducing extraneous COGS.


The proposed system auto-tunes its hyper-parameters to balance between performance and operational cost with minimal to no engineering input. 
Evaluated using large-scale production data, \sysname achieves up to 43\% reduction in cluster idle time compared to static pooling when targeting 99\% pool hit rate. Currently deployed in production, \sysname is on track to save tens of million dollars in COGS per year as compared to traditional pre-provisioned pools.

\end{abstract}

\maketitle

\pagestyle{\vldbpagestyle}
\begingroup\small\noindent\raggedright\textbf{PVLDB Reference Format:}\\
\vldbauthors. \vldbtitle. PVLDB, \vldbvolume(\vldbissue): \vldbpages, \vldbyear.\\
\href{https://doi.org/\vldbdoi}{doi:\vldbdoi}
\endgroup
\begingroup
\renewcommand\thefootnote{}\footnote{\noindent
	This work is licensed under the Creative Commons BY-NC-ND 4.0 International License. Visit \url{https://creativecommons.org/licenses/by-nc-nd/4.0/} to view a copy of this license. For any use beyond those covered by this license, obtain permission by emailing \href{mailto:info@vldb.org}{info@vldb.org}. Copyright is held by the owner/author(s). Publication rights licensed to the VLDB Endowment. \\
	\raggedright Proceedings of the VLDB Endowment, Vol. \vldbvolume, No. \vldbissue\ %
	ISSN 2150-8097. \\
	\href{https://doi.org/\vldbdoi}{doi:\vldbdoi} \\
}\addtocounter{footnote}{-1}\endgroup

\section{Introduction}
\label{sec:intro}

Cloud computing has emerged as a top choice for executing big data analytic workloads in various business domains.
To cope with rapidly increasing demand, cloud vendors (e.g., Amazon AWS~\cite{aws}, Microsoft Azure~\cite{azure} and Google GCP~\cite{gcp}) have funneled sizable resources into their own managed Spark services~\cite{zaharia2010spark,nghiem2016towards}, including Google Cloud's Serverless Spark~\cite{spark1}, Spark through Vertex AI~\cite{spark3}, Azure HDInsight~\cite{azure1}, Azure Synapse Analytics~\cite{azure2} and AWS EMR~\cite{aws1}. 
The flexibility of such cloud offerings allows users to easily lease and release compute resources as required and, consequently, enjoy potentially significant cost-effectiveness. However, to provide such flexibility, service providers must address various challenges with respect to resource provisioning.


The implementation of multi-tenancy and scalability in such systems results in prolonged latencies in accessing clusters.
With Azure Synapse~\cite{azure2}, it is common to experience a cluster initialization time of over 60 seconds. According to Databricks~\cite{databricks}, this provisioning time can be even longer than the duration of the job execution.
However, proactive \edit{provisioning} solutions are often challenging due to: (1) The unpredictability of user behavior and, 
(2) the difficulty in developing any policy that both enhances performance and decreases cost-of-goods-sold (COGS),  
which requires an explainable, comprehensive decision-making process in real-world production. 
Accurately modeling multi-tenant cloud performance and its impact on customer experience can be complicated involving complex modeling
~\cite{chen2019cost,zhu2021kea}. 

\subsubsection*{\textbf{State-of-the-art approach}}
Proactive auto-scaling \edit{(after application starts) has been introduced in data stream processing engines (e.g., Apache Storm~\cite{storm}, Apache Flink~\cite{flink}) and network provisioning~\cite{bibal2019rlpas,zhang2017proactive} to dynamically \textit{scaling up} (or down) the compute resources} when the workload is heavy~\cite{lee2017framework}, and performance modeling is developed to ensure the QoS requirements are fulfilled ~\cite{liu2014aggressive}. Time-series forecasting is used to determine the possible repeating patterns as inputs~\cite{baldan2016forecasting}. A detailed review of similar applications in streaming systems can be seen in~\cite{lorido2014review}. 

For Spark~\cite{zaharia2016apache} clusters, there has not been an automated solution to manage cluster provisioning. Some vendors, like Databricks~\cite{databricks}, provide mechanisms to maintain clusters until a fixed threshold of idle time is reached and offer customers instruction on ``best practices'' for managing Spark clusters on their own. Despite its potential benefits, this approach still requires manual tuning from the user's perspective and may result in unsatisfactory performance.

In this paper, we tackle the issue of improving the customer experience of waiting for the \textit{initialization} of Spark clusters where a cluster needs to be prepared for a newly submitted Spark job \edit{at its startup time}, which is one of the major bottlenecks for many Spark systems. 
\edit{Compared to auto-scaling while the application is running, this problem can be more challenging because at the application submission time or even before (if supporting proactive provisioning), there can be little-to-no information known for a particular customer or application. Auto-scaling relies on real-time information such as cardinality estimates, number of tasks queued, etc., which becomes available when the application starts, to adjust the number of nodes and executors.}
Auto-scaling during the lifetime of an application is out of scope for this paper, and \edit{for \fabric, }there is a different service to scale up/down the number of nodes in the compute cluster as well as the number of executors in real time based on the incoming workload characteristics and the tracking of task execution, which include richer information about the application and the Service Level Agreements (SLAs) that need to be met~\cite{thonglek2021auto,baresi2018towards,oh2016sla}.

\subsubsection*{\textbf{Challenges}}
In order to reduce the wait time for cluster initialization, we aim to use machine learning to proactively provision Spark clusters. However, this presents a number of challenges, including::

\mypar{Uncertainty of user behavior [C1]} 
Intuitively, it would be possible to provision a cluster in advance if we could accurately predict when a customer will submit a job. However, this is difficult to achieve in practice due to the high degree of uncertainty of individual user behavior. 
\edit{Training an individual model for each customer, as is done in ~\cite{poppe2020seagull}, is not feasible due to scalability constraints}.

\mypar{Difficulty of modeling performance-cost trade-offs [C2]}
In tandem with the proactive provisioning mechanism, one needs to model the performance observed by customers (for example the wait time for accessing a cluster and starting a job) and estimate the extraneous COGS from the operator's point of view. Any mismanagement of resources, including over-provisioning or under-provisioning, will result in either significant financial losses or an unsatisfactory customer experience. 

\edit{\mypar{Compliance for service level agreements [C3]} The} inherent unpredictability and opacity of machine learning is always the biggest concern. While most cloud operators are mandatory to meet specific service level agreements, a robust and consistent algorithm is critical. \edit{However, the algorithm may fail to converge in certain corner cases, and using a black-box approach like ML introduces significant challenges in debugging and error triage.}

\mypar{Requirements of full automation [C4] and low latency [C5]} For a production-level system, it is necessary to have a fully automated and reliable solution.
Additionally, the provisioning system must be able to adapt to a constantly changing environment by taking into account the real-time state of the system. 
To achieve this, it is necessary to develop and maintain a low-latency monitoring system, as well as simple and efficient algorithms.

\subsubsection*{\textbf{Introduction to \sysname}}
To overcome these challenges, we propose \sysname, a self-adaptive solution that proactively creates clusters based on monitoring of demand.


\textit{We introduce the notion of a Spark ``live pool'', where a number of clusters are proactively created and pre-configured for various users. [C1].}
Whenever a customer requests a cluster, one cluster will be immediately evicted from the pool and made available for use. 
At the same time, \edit{a new cluster is provisioned and added to maintain a constant cluster number, a process referred to as "re-hydration." }



\textit{We introduce a self-tuning system to dynamically learn the optimal pool size based on demand, considering the cost-performance trade-offs [C2, C5].} 
One of the biggest concerns of the live pool mechanism is the COGS. At the scale of Microsoft Fabric~\cite{fabric}, we expect simple threshold-based provisioning to quickly exceed 10,000 CPU cores, resulting in tens of millions of dollars in COGS. 
To address this challenge, our work proposes dynamically adjusting the size of the pool based on customer demand. 
We propose an efficient linear programming (LP) solution to model the two factors (performance and cost) based on the Pareto frontier to determine the optimal pool size. 
A self-adaptive system is proposed that automatically balances the trade-off between the cost of maintaining idle clusters and the potential for long wait times for customers.

\textit{We introduce an efficient, and robust hybrid time-series forecasting algorithm of the future demand at the aggregate level with high accuracy and robustness [C1, C3, C5].} The models predict future demand, measured by the cluster request rate, based on historical demand and the latest observation of the cluster creation request rate. The predictions then serve as inputs to the optimization model. With an end-to-end run time (training, inferencing, and optimizing) reduced to \textit{mere seconds}, we ensure that the recommendations are always up-to-date by retraining the model with high frequency (e.g., < 5min). To improve the robustness of the model prediction [C3], we developed a new policy to smooth the input data, which significantly reduces performance regressions.

\textit{We have implemented a real-time monitoring system that provides continuous inputs to constantly learn the optimal provisioning policy [C4, C5].} 
\edit{The telemetry data collected is then fed into our optimization algorithm in real-time}. The same dashboard is also used to evaluate the performance of the system.

\textit{The end-to-end solution is \edit{lightweight and implemented in C\# as an integral part of the core Spark infrastructure [C5].}} The recommendation engine can be executed in a single invocation of the pipeline, with fast and accurate recommendation generation and persistence in configuration files in mere seconds, ready to use for the pooling service.
We integrate the modules with the new Fabric service, the new \msft data analytics offering that supports Spark~\cite{fabric}, deployed in all production Azure regions \edit{in Nov, 2023}.

\subsubsection*{\textbf{Contribution}}
In sum, our contributions are:
\begin{itemize}
    \item A simple linear programming formulation for solving the optimal pool size that captures the trade-off between improving performance and reducing COGS;
    \item An efficient hybrid ML algorithm combining deep learning and traditional ML for predicting future demand and optimal pool size with extremely low latency;
    \item A robust strategy to account for demand uncertainty to ensure high service level;
    \item Deployed in production, we \edit{achieved millions of dollars in annual COGS} compared to preexisting pooling.
\end{itemize}


The remaining sections are organized as follows:
Section~\ref{sec:bg} provides background and motivates the problem.
Section~\ref{sec:overview} presents the overall design principles and architecture. 
Sections~\ref{sec:Saa} and~\ref{sec:ml} describe the optimization and ML modules, respectively.
Section~\ref{sec:exp} goes over how we evaluated the algorithm using production data.
Section~\ref{sec:related} discusses related work, and Section~\ref{sec:conclusion} concludes the paper.

\section{Background}
\label{sec:bg}


One common issue related to Spark cluster initialization is a long wait time~\citep{databricks}.
Complexity in the underlying infrastructure introduces a wide range of potential slowdowns that are difficult to detect, diagnose and mitigate. 
Examples of this complexity include hardware heterogeneity, unreliable network communication, and inter-node service coordination which is compounded by the strict multi-tenancy requirements in the cloud.
Efforts are undergoing to reduce the \textit{tail latency} of cluster initialization time, including making hedged requests~\cite{lu2014mechanisms}
and using tied requests \citep{dean2013tail,lu2014mechanisms}.
However, these approaches are not able to completely address the issue. 

\edit{
In May 2023, \companyname announced \fabric, a new data analytics that offers both data engineering and data science experiences, operating as a multi-tenant managed Spark service, with a security boundary scoped to individual users.
}
\al{As a reader, I might wonder, why can't the existing infrastructure be improved? Maybe we can discuss briefly the limitations of the current system?}\yz{I think that would be great. Alex could you add some?}\al{Synced offline on this. One option here can be to add references to other papers that go into some of the infra limitations related to Spark. Another option could be to provide some type of general statement.}
On this platform, the typical underlying process to initiate a Spark session consists of ~60-120 seconds for the cluster creation and ~30-40 seconds for the session creation~\citep{zaharia2010spark,aleksiyants2015implementing}. When Generic Job Service requests a new cluster, 
\edit{
This prolonged process is primarily influenced by four main processes: VM configuration, allocation, and boosting time; stitching VMs to form Spark clusters; configuring libraries; and creating a Spark session—each taking 30-60 seconds, contributing to the extended wait time.}

To minimize the latency experienced by the end-user, we propose to proactively provision Spark clusters.
\edit{In this work}, we propose to create a shared pool of actively-running clusters, which we call a \textit{live pool}. We categorize live pools into two buckets: \textit{session pools} and \textit{cluster pools}, i.e. interactive and batch mode respectively. Both consist of pooled clusters; the difference is that session pools also have an actively-running \textit{Spark session} in each cluster, which we call a \textit{pooled session}. Session pools are useful for notebook scenarios, when a pre-created session can be used to run a notebook instantaneously. Pooled clusters, by contrast, are useful for running batch jobs with pre-defined job definitions (e.g., a json file that describes a .jar file location, Spark configurations, etc.) and Spark sessions that require ad hoc customization. For the rest of the paper, we discuss the methodology with respect to cluster pools, though the same can be applied to session pools.
Similar to the concept of inventory management in retailing~\citep{dehoratius2008retail}, we maintain a constant number of resources in a given pool, and upon receiving a client request, a pre-provisioned resource can be used instantly. To maintain the target number of resources in the pool, we send a new request, referred to as a re-hydration request, to Generic Job Service to add a new cluster or session back to the pool whenever a pooled resource is consumed or fails (due to exceeding a pre-defined lifespan or unexpected system failures). \edit{For \fabric, two pools per region (one for session and one for cluster) with a fixed cluster size, e.g., 3-median nodes, are created. }

\begin{figure}[t]
    \includegraphics[width=0.4\columnwidth]{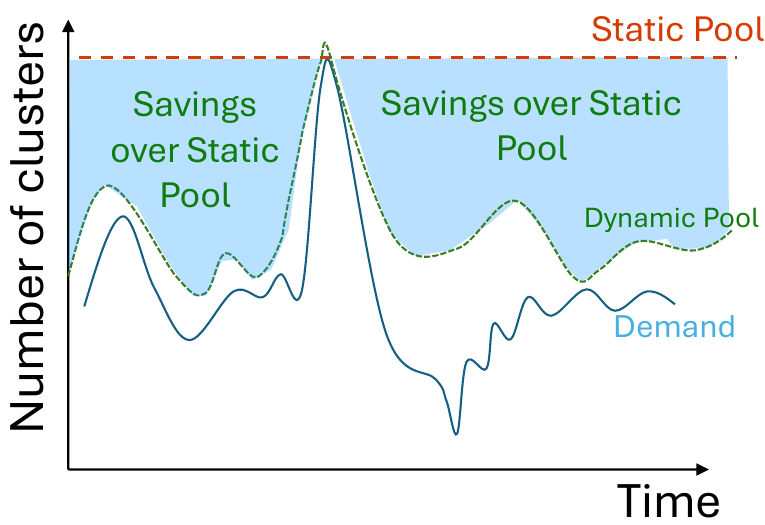}
    \vspace{-0.3cm}
    \caption{\edit{Benefits of \sysname}}
    \label{fig:benefits}
    \vspace{-0.6cm}
\end{figure}

The general idea of \sysname is to dynamically determine the optimal number of resources in a pool and scale the pool up or down as needed in real-time. A larger pool can lead to wasted COGS in a low-demand scenario. On the other hand, a smaller pool has a higher likelihood of being drained out in high-demand scenarios, where numerous customers need clusters or sessions at the same time and the system does not have enough time to sufficiently replenish the pool. The client request in this situation must go through the original protracted startup process (referred to as ``on-demand"). \edit{With dynamic pooling, by adjusting the pool size according to (predicted) demand, we can achieve potentially significant savings over the static pool (see Figure~\ref{fig:benefits}).}
Given that ML algorithms are in general never perfect, with margins in prediction errors, the optimal provisioning strategy remains a challenge.

\section{Intelligent Pooling Overview} \label{sec:overview}

\begin{figure}[t]
    \includegraphics[width=0.82\columnwidth]{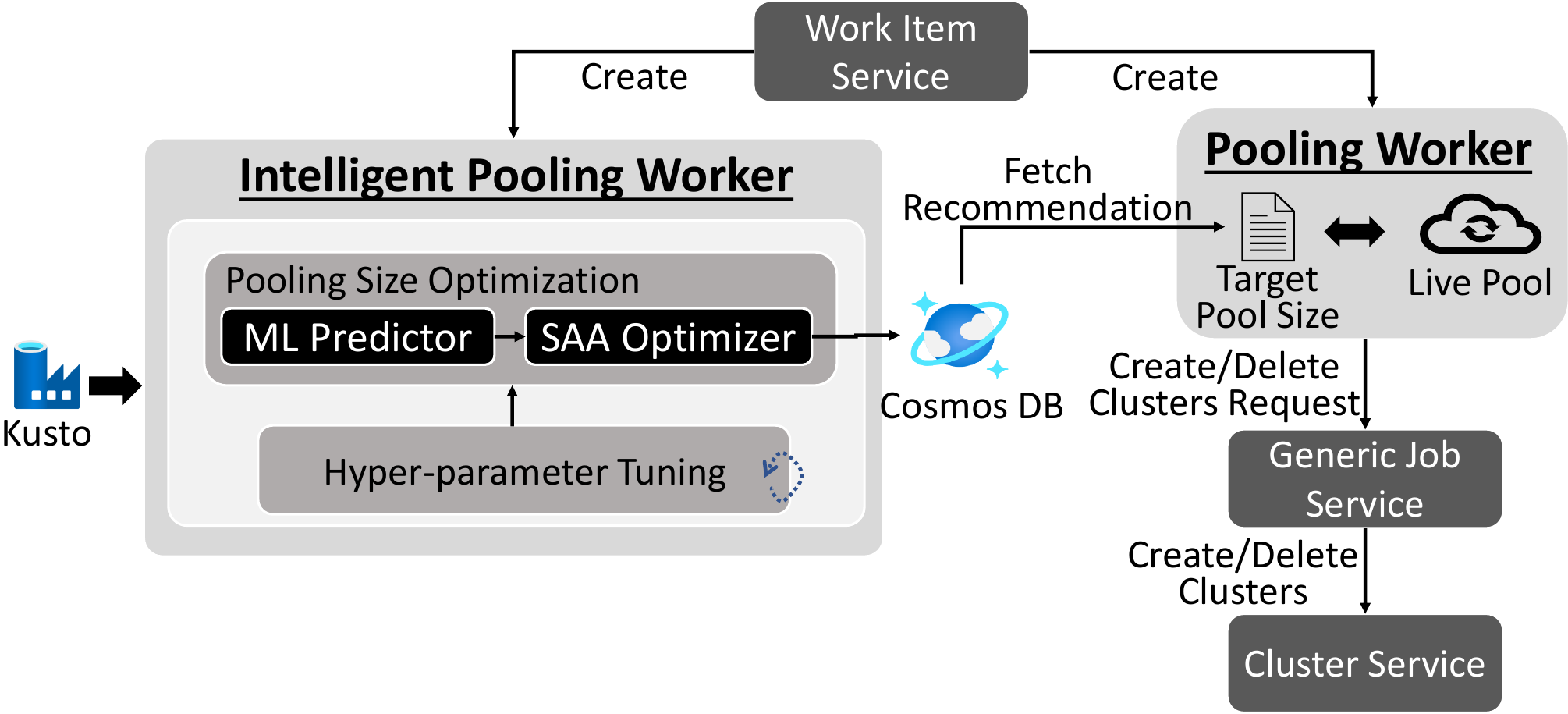}
    \vspace{-0.3cm}
    \caption{Architecture}
    \label{fig:architecture}
    \vspace{-0.6cm}
\end{figure}

In this section, we discuss the overall architecture of \sysname.
\al{It might be worth including Cluster Service in Figure 1. Since I added some references to it, it would be good for the reader to be able to visualize how Generic Job Service interacts with it.}
\sysname consists of two main modules (see Figure~\ref{fig:architecture}):
\begin{itemize}
    \item The \textbf{Sample Average Approximation (SAA) Optimizer} formulates a simple linear programming problem to optimize the pool size based on input demand, which can either be historic or
    predicted by the ML Predictor. The optimized results are then saved as configuration files in Cosmos DB~\cite{cosmosdb} (Section~\ref{sec:Saa}). 
    \item The \textbf{ML Predictor} makes real-time time-series predictions
    by constantly fetching historic observations from the Kusto store~\cite{kusto}. 
    Using predicted demand as opposed to historic demand helps to react more accurately to real-time changes in the system, though it can potentially lead to longer model-training latencies (Section~\ref{sec:ml}).
\end{itemize}

\sysname leverages the existing infrastructure used by live pools for its own execution and deployment. Specifically, 
\begin{itemize}
    \item \textbf{Generic Job Service} is responsible for orchestrating Spark batch jobs and interactive sessions; providing APIs to perform CRUD operations on Spark jobs; and 
    managing and processing Spark job-related metadata.
    \item \textbf{Cluster Service} is responsible for requesting virtual machines (VMs) from Azure and "stitching" them to form Spark clusters; providing APIs to perform CRUD operations on Spark clusters; and managing and processing Spark cluster-related metadata.
    \item \textbf{Work Item Service} is a background service that supports various workloads, including but not limited to Spark cluster and session pooling, Spark job submissions, and the \sysname infrastructure. It is responsible for monitoring available \textit{work items} that represent these workloads and spinning up worker processes that execute the workloads.
    \item The \textbf{Intelligent Pooling Worker} is responsible for periodically running the ML pipeline that includes the aforementioned SAA Optimizer and ML Predictor and persisting the recommendation files in Cosmos DB~\cite{cosmosdb}.
    \item The \textbf{Pooling Worker} is responsible for maintaining a target pool size by invoking Generic Job Service to create and delete resources and, if applicable, fetching the latest pool size recommendation file emitted by an Intelligent Pooling Worker.
\end{itemize}



The hyper-parameter tuning module is developed to constantly fine-tune the hyper-parameters for the optimization algorithm to avoid over- or under-allocating resources by setting the tuning knobs to minimize the COGS while satisfying the SLA. 
It can be executed at a lower frequency while the ML pipeline runs at a higher frequency such that it captures the rapid change of the environment and adapts faster to the demand. More details on the hyper-parameter tuning process can be seen in Section~\ref{sec:tuning}.

\section{Sample Average Approximation Optimizer}\label{sec:Saa}
In this section, we formulate the live pool mechanism as a queuing system and propose linear programming to reach optimality.

\subsection{The Live Pool Mechanism}

\begin{figure}[t]
    \centering
    \includegraphics[scale=0.35]{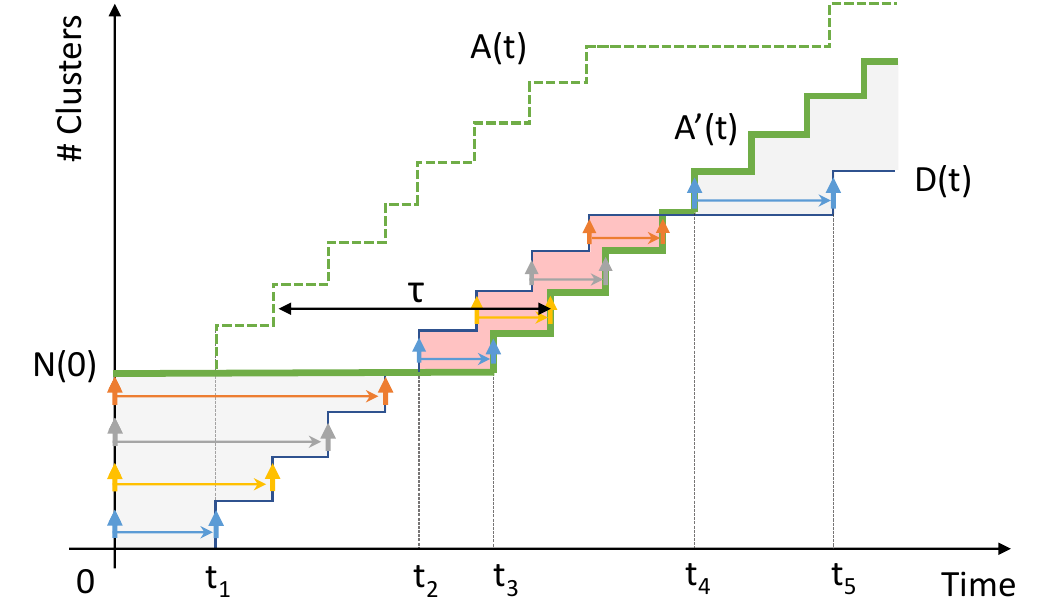}
    \vspace{-0.2cm}
    \caption{Cumulative cluster creation number $D(t)$, cumulative re-hydration requests $A(t)$, 
    number of clusters ready $A'(t)$ at time $t$, and pool size at time 0, $N(0)$, the total wait time (red area) for obtaining a running cluster and idle time (grey area) for clusters in the pool.
    }
    \label{fig:total_wait_time} \label{fig:cumulative}
    \vspace{-0.6cm}
\end{figure}

We illustrate the live pool mechanism through a graph plotting a set of cumulative values (see Figure~\ref{fig:cumulative}). 
Specifically,

\begin{tabular}[H]{lp{6.9cm}}
$D(t)$: &the cumulative number of clusters requested by customers (demand);\\
$N(t)$: &the target pool size as a function of time that we want to maintain;\\
$A(t)$: &the cumulative number of cluster re-hydration requests made to add a cluster to the pool in order to keep it at the target pool size of $N(t)$;
\\
$\tau$: &the cluster initialization time, i.e., the time lag before a cluster can be ready for use after the creation request is sent;\\
$A'(t)$: &the cumulative number of clusters ready for use.\\
\end{tabular}


For instance, at time $t = 0$, 
a pool is created with $N(0)=4$ clusters, and whenever a user request for a new cluster is received, (corresponding to an increase in $D(t)$, e.g., at $t_1$), a cluster will be emitted from the pool to be used by the customer. At the same time, the pooling worker will initiate a re-hydration request to Generic Job Service to add a new cluster back to the pool (corresponding to an increase in $A(t)$).
As a result, the curve of $A(t)$ can be seen as a simple ``shift-up" of the curve of $D(t)$, and the gap between them equals the target pool size $N(t)$.
Similarly, because cluster creation takes time, the cumulative number of clusters that are ready to use, $A’(t)$, is a ``shift-right" of the number of requests made, $A(t)$, by the cluster creation latency, $\tau$:
\begin{align}
A(t)&= D(t) + N(t),\quad \forall t \label{eq:cons1}\\
A'(t)&= A(t-\tau),\quad \forall t\geq \tau \label{eq:cons2}\\
A'(t)&= N(0),\quad \forall t < \tau. \label{eq:cons3}
\end{align}

Assuming a first-come-first-serve rule, the clusters in the pool will be acquired by customers based on the user request arrival time. For instance, in Figure~\ref{fig:cumulative}, the first four clusters in the pool will be used by the first four requests that are received (blue, yellow, grey and orange respectively). The fifth created cluster is triggered by the arrival of the blue request at $t_1$, and is not ready until $t_3$. It will be used by the fifth request received at $t_2$.

\subsection{Optimal Pool Size}
There are two factors to optimize for: 
\begin{itemize}
    \item The total idle time pooled clusters are alive and unused by customers; and
          \item The total wait time for customers (when the pool is drained out and the customer must wait for the full cluster startup duration).
\end{itemize}


In Figure \ref{fig:total_wait_time} we show the one-to-one mapping between a created cluster and the request that will use the corresponding cluster based on a first-come-first-serve (FCFS) rule. We observe that whenever $A'(t)>D(t)$, idle time occurs, and whenever $A'(t)<D(t)$, wait time occurs. For example, at time $t=0$, four clusters have been created, and they will be used by the first four requests that come in (blue, yellow, grey and orange). For those clusters, their aggregate idle time equals the area of the gap between the curves of $A'(t)$ and $D(t)$, highlighted in grey. However, the fifth cluster creation request on curve $A(t)$ (marked in blue) was not ready until $t_3$ while the fifth customer request on curve $D(t)$ occurs at $t_2$. Therefore, for the fifth request, the customer has to wait for $t_3-t_2$, and the wait time was highlighted in red (similar for the sixth, seventh and eighth requests in yellow, grey, and orange respectively)\footnote{Note that currently, when a pool is drained out, ``on-demand'' cluster creation requests will be sent to accommodate the cluster requests, and their wait time becomes $\tau$. The clusters being ready at $t_3$ and onwards will be served to later requests. In this case, the FCFS rule is violated as the order of using the clusters is modified. In this work, we still assume FCFS for simplicity as an approximation of this mechanism.}. 

In sum, the total wait time of customers is the red area where $A'(t)<D(t)$, and the total idle time of clusters in the pool is the grey area where $A'(t)>D(t)$. With this, we can estimate the optimal pool size where both the total wait time and the idle time are minimized. Specifically, we can leverage linear programming with minimization as the objective to calculate the areas:
\begin{align}
    \Delta^{+}(t) &\geq A'(t) - D(t),\quad \forall t\label{eq:cons4}\\
    \Delta^{+}(t) &\geq 0\quad, \forall t\\
    \Delta^{-}(t) &\geq D(t) - A'(t),\quad\forall t\\
    \Delta^{-}(t) &\geq 0,\quad \forall t. \label{eq:cons7}
\end{align}

If the objective function involves minimizing $\Delta^{+}(t)$ and $\Delta^{-}(t)$, one can prove that in the optimal solution, if $A'(t) \geq D(t)$, $\Delta^{+}(t) = A'(t) - D(t)$ and $\Delta^{-}(t)=0$. If $A'(t) \leq D(t)$, $\Delta^{-}(t) = D(t) - A'(t)$ and $\Delta^{+}(t)=0$. And the sum of $\Delta^{+}$ and $\Delta^{-}$ calculates the total area of grey (idle time) and red (wait time) respectively. And the constraints are all linear.


\subsubsection*{\textbf{Sample Average Approximation (SAA) Optimizer}}
Based on the above discussion, we can formulate the optimization program with the objective to minimize the total cost (considered as a weighted sum of the wait time and idle time). We use the sample average approximation (SAA) method~\cite{kleywegt2002sample} based on the input demand data to minimize the expected total cost over the whole observed period. 
Denote $\alpha$ and $\beta$ the hyper-parameters representing the penalty of having long idle time versus wait time, a larger $\alpha$ will result in an optimal solution trying to minimize the idle time more than wait time, and vice versa. By changing the hyper-parameter values, one can achieve the full Pareto curve~\cite{floratou2017dhalion, pareto} of the trade-off between idle time and wait time.
The detailed formulation is as follows:
\begin{align}
&\min  \alpha \cdot\sum_{t}{\Delta^{+}(t)} + \beta \cdot \sum_{t}{\Delta^{-}(t)} \label{eq:op1}\\
&~\text{s.t.},\quad\quad\quad\quad\quad~\eqref{eq:cons1} - \eqref{eq:cons7}. \nonumber
\qedhere
\end{align}

With this minimization formulation, for the optimal solution, $\Delta^{+}(t)$ equals the number of idle clusters at time $t$, and $\Delta^{-}$ the queued demand. All the constraints are linear.
For the maximum number of requests made, one can add a constraint with the maximum number of requests per time interval, MAX NEW REQUEST:
\begin{align}
    N(t) - N(t-1)\leq \text{MAX NEW REQUEST} \quad \forall t\ge1.
\end{align}
For the analyzed system, we also added the following constraints:
\begin{align}
   \text{MIN POOL SIZE}\leq N(t)\leq \text{MAX POOL SIZE} \quad \forall t\ge1,\label{eq:minmax}\\
   N(t) = N(\nint{t / \text{STABLENESS}} * \text{STABLENESS}) \quad \forall t\ge1.\label{eq:stable}
\end{align}
where Constraint~\eqref{eq:minmax} sets the minimum and maximum pool size (i.e., MIN POOL SIZE and MAX POOL SIZE), and Constraints~\eqref{eq:stable} ensures that the pool size is stable for $\Delta t=\text{STABLENESS}$ intervals. \edit{In production, the MIN POOL SIZE and MAX POOL SIZE are set according to regional capacity.}

For a simplified intelligent pooling policy, one can also add constraints to ensure that the pool size for the same day of week or time of day is the same as for a more static controlling policy. Note that all the constraints are still linear and can be solved by commercial solvers with low latency (in a few seconds).



\section{ML Predictor}\label{sec:ml}
In this section, we detail the prediction model and pipeline that we chose. Specifically, Section~\ref{sec:mlinit} discusses initial model exploration and Section~\ref{sec:mllimit} discusses the limitations. Section~\ref{sec:mlhybrid} proposes a hybrid model that combines a traditional machine learning algorithm with deep models. Section~\ref{sec:ml2step} proposes an alternative way to combine the SSA optimizer with the ML predictor. 

\subsection{Cost-Performance Trade-offs}\label{sec:mlinit}
We evaluate the following four different models, each representing a different category/approach: (1) Singular Spectrum Analysis (SSA) \cite{golyandina2014basic} implemented by ML.NET~\cite{ahmed2019machine}; (2) Inception Time \cite{ismail2020inceptiontime}; (3) TST \cite{zerveas2021transformer}; and (4) mWDN \cite{wang2018multilevel}.
\edit{These models are chosen since each represents a different category---SSA is a traditional ML model, TST is a transformer-based deep learning approach, mWDN is wavelet decomposition-based approach and Inception Time is a 1D convolution model.} 

\edit{To train the models, we use an 80-20 train-test split. Specifically, for the deep learning models, the training set is further split into a 90-10 train-validation set. For DNN models, we} use the validation set to ensure we do not overfit to the training set and to trigger an early stop. To directly embed the estimation of the wait-idle time trade-off into the training process, 
we use a modified loss function similar to the estimation of $\Delta^{+}$ and $\Delta^{-}$ as in Equations~\ref{eq:cons4}-\ref{eq:cons7}, which is a proxy of the true wait and idle time to capture the trade-offs between cost and performance.
We define the loss function $\mathcal{L}$ as:
\begin{align}
\mathcal{L} & = \alpha' \cdot \delta^{+} + (1 - \alpha') \cdot \delta^{-} \label{eqn:loss}\\
\delta &= y - \hat{y}  \\ 
\delta^{+} & = \delta : \delta > 0 \\
\delta^{-} & = -\delta : \delta < 0 
\end{align}

where $y$ is the ground truth time series, $\hat{y}$ is the predicted output and $\alpha'$ is the hyper-parameter that controls the relative importance of idle time and wait time during optimization and training.   We observe that the modified loss function in Equation \eqref{eqn:loss} allows for the models to perform better at the extremes of the wait time and idle time constraints. It also allows the model to estimate the demand at a higher/lower percentile, tailored to our business needs.

\subsection{Limitations} \label{sec:mllimit}
From our experiments,
we found that the rate at which we update the model has a big impact on the idle time (i.e., cost savings). Thus, it was critical to identify and deploy models that were fast to update. \edit{We found that deep models (mWDN, TST and Inception Time) were significantly slower to train compared to SSA (see Figure \ref{fig:scaling} on data scaling) though they offer flexibility with customized loss functions, allowing tailoring to different performance and cost preferences for demand prediction—whether conservative
or aggressive. Lacking such flexibility, SSA failed to achieve sufficiently low wait times (refer to Figure \ref{fig:trade_off_small_uswest2_idle_vs_wait}). To address these two limitations, we propose a new \textit{hybrid model} which combines the best of both.}

\subsection{\edit{Hybrid Model: SSA+}}\label{sec:mlhybrid}
To address the above-mentioned limitation, in this paper, we propose the hybrid model to achieve low training latency and relatively good trade-off performance. We address the issues with SSA and the deep models by combining certain parts of each. \edit{The reason SSA fails to achieve low wait times is because there is no way to specify and control how much the predicted request rate must overshoot the ground truth (see Section~\ref{sec:expppl}). If the predicted usage/pool size is larger than ground truth, this will result in a larger pool size, lowering the average wait time. With deep models, the overshoot is controlled using the loss function defined by Equation \eqref{eqn:loss}. However, the issue with deep models is that the models are too computation-intensive for the task at hand and need lots of data and computational resources to train the over-parameterized model. Thus, the proposed hybrid model consists of an SSA forecaster followed by a shallow two-layer neural net ($\approx$ 30 parameters, with ReLU activation for non-linearity) which acts as an error predictor. This error predictor can be trained using the loss from Equation \eqref{eqn:loss} to learn the overshoot or undershoot needed to achieve the target wait time.} The improved trade-off performance of the hybrid model can be seen in Figure~\ref{fig:trade_off_small_uswest2_idle_vs_wait}.

\subsection{End-to-end Recommendation Engines}\label{sec:ml2step}
While modern ML approaches can achieve high performance, they are not immune to errors. Thus, any processing that is performed post-ML prediction must not introduce more errors by propagation. This is similar to the game of telephone, where the ML model makes a prediction and subsequent processing introduces enough uncertainty to make the prediction useless. With the goal of minimizing the potential for errors, we explored two end-to-end recommendation pipelines:
\begin{itemize}
    \item \textbf{2-step} pipeline where the ML model is trained on the input cluster request rate data. The ML model predicts cluster request data, which is then fed to the SAA optimizer which outputs the predicted optimal pool size.
    \item \textbf{End-to-End (E2E)} pipeline where we apply the SAA optimizer on the historic data, providing a ground truth optimal pool size for the past. The historic optimal pool size is then used to train the ML model, which predicts the optimal pool size for the future.
\end{itemize}
From our experiments, we found 2-step pipelines have a better Pareto curve when targeting low wait times (Section~\ref{sec:exp}).
\section{Auto Tuning Parameters}
\label{sec:tuning}
Constantly adapting the hyper-parameters to meet business needs is a challenge and adds to the cost of maintaining the service. Particularly, in production we want to achieve our service-level agreement on the performance (wait time) while minimizing the idle versus wait time penalty. Thus, we propose using a self-adaptive hyper-parameter tuning mechanism to close the feedback loop. We constantly monitor the system behavior (pool hit and pool misses) and adjust for the parameters accordingly such that we can always maintain the optimal balance as desired.

To achieve this, we eliminate the $\beta$ hyper-parameter from Equation \ref{eq:op1} and rewrite the objective function as:
\begin{align}
&\min  \alpha' \cdot\sum_{t}{\Delta^{+}(t)} + (1-\alpha') \cdot \sum_{t}{\Delta^{-}(t)}, \label{eq:ssa_min_no_beta}
\end{align}
where $0 \le \alpha' \le 1$.
It is easy to show that the formulation is equivalent to Equation~\eqref{eq:op1}.

Thus we have only one hyper-parameter to tune, which reduces our search space. We can model the relation between the business requirement (customer wait time) $t_{\text{wait}}$ and the hyper-parameter $\alpha'$. We approximate the relation $\alpha' = f(t_{\text{wait}})$ to be piece-wise linear. With this approximation, we try to fit the best line based on the previous 10 data points and update the value iteratively.

\section{Experiment}
\label{sec:exp}
In this section, we present the results of applying SAA on historic data (Section~\ref{sec:expsaa}), compare the performance of various ML models (Section~\ref{sec:ml_model_comp}) and compare the performance of the two end-to-end pipelines integrated with the optimizer by displaying their wait-idle time trade-off curves (Section~\
\ref{sec:expppl}). We discuss the model efficiency in Section~\ref{sec:scaling} and deployment results in Section~\ref{sec:deployment}.

\edit{In all assessments, we consolidate the input time series data into 30-second intervals, where the data signifies the number of cluster requests per interval. We maintain a constant pool size for 5 minutes to prevent abrupt changes in size recommendations. Various combinations of penalty values, such as $\alpha$, $\beta$, and $\alpha'$, were examined to achieve diverse trade-offs between perf and cost. The evaluation was conducted on a node with 6 vCores and 64 GB RAM.}

\subsection{Sample Average Approximation (SAA) optimizer}\label{sec:expsaa}
\begin{figure}[t]
    \includegraphics[width=0.93\columnwidth]{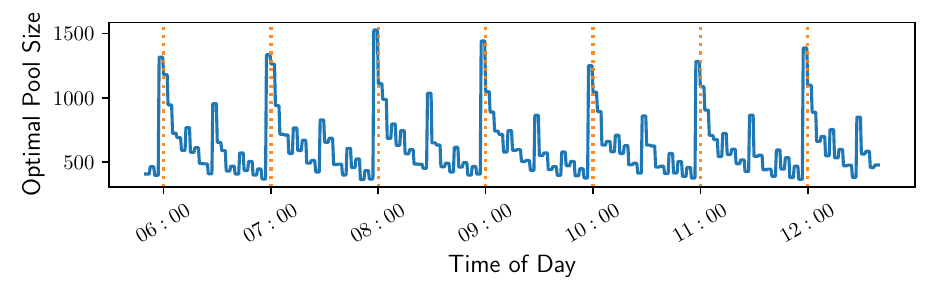}
        \vspace{-0.6cm}
    \caption{Pool size increases advance demand.}
        \vspace{-0.6cm}
    \label{fig:poolsizeadvance}
    \end{figure}

We extracted production Azure Synapse data in the East US region from July 01 to July 15 in 2022 with hundreds of thousands of cluster requests. 
\edit{We} estimated the optimal pool size by time of day and type of day (weekday versus weekend) using historic data. Several interesting findings emerged:

\textit{The pool size is correlated with the number of requests.} In Figure \ref{fig:poolsizeadvance}, we find that the pool size increases 5 minutes before the start of every hour that is, 5:55, 6:55, 7:55, etc. This is due to the fact that many jobs are scheduled at 6AM, 7AM, etc. The optimization proactively prepares for this surge by increasing the pool size to cope with this demand.
        
\textit{There is a trade-off between longer wait time and longer idle time.} As discussed previously, a larger pool size generally results in longer idle times and a decrease in the likelihood of the pool being drained out. We can tune the value of the cost penalty in the objective function to tune the pool size and obtain a Pareto curve. 

\textit{High frequency in which the pool size is updated can improve both COGS and performance}. 
We observed that by decreasing the $STABLENESS$ as in Equation~\eqref{eq:stable} the Pareto curve shifts towards the lower left, indicating better perf-cost trade-offs.
However, in production, we are not able to update the pool size too frequently and potentially decreasing the pool size will also result in cancellation of re-hydration requests.


\subsection{ML Model Comparison}
\label{sec:ml_model_comp}

\edit{
\begin{table}[h]
\caption{\edit{Performance comparison using a 2-step pipeline. }}
\vspace{-0.2cm}
\begin{adjustbox}{width=\columnwidth}
\begin{tabular}{|c|c|ccccc|}
\hline
\multirow{2}{*}{Region} & \multicolumn{1}{c|}{\multirow{2}{*}{Node Size}} & \multicolumn{5}{c|}{MAE $\downarrow$}              \\ \cline{3-7}
                        & \multicolumn{1}{c|}{}                             & SSA+  & SSA\cite{golyandina2014basic}   & mWDN\cite{wang2018multilevel}  & TST\cite{zerveas2021transformer}  & IncpT\cite{ismail2020inceptiontime} \\ \hline
West US 2               & Small                                            & \textit{\textbf{9.54}} & 14.13 & 10.28 & 10.86 & 10.00 \\
East US 2               & Small                                            & 7.78 & 8.02  & 7.23  & \textit{\textbf{7.13}}  & 7.64  \\
West US 2               & Medium                                           & 5.54 & 4.82  & \textit{\textbf{3.77}}  & 4.28  & 4.01  \\
East US 2               & Medium                                           & 1.42 & 1.60  & \textit{\textbf{1.26}}  & 1.33  & 1.33  \\
West US 2               & Large                                            & 3.28 & 3.67  & \textit{\textbf{3.21}}  & 3.23  & 3.40  \\
East US 2               & Large                                            & 1.89 & 2.42  & \textit{\textbf{1.79}}  & 1.90  & 1.98  \\ \hline
\multicolumn{2}{|c|}{Average}                                                    & 4.91 & 5.78  & \textit{\textbf{4.59}}  & 4.79  & 4.73    
\\ \hline
\end{tabular}
\end{adjustbox}
\label{tab:perf_comp}
\end{table}
}

To identify the best ML model, we compared the performance of Singular Spectrum Analysis (SSA) \cite{golyandina2014basic}, Inception Time \cite{ismail2020inceptiontime}, TST \cite{zerveas2021transformer} and mWDN \cite{wang2018multilevel} models \edit{with our proposed hybrid model (SSA+) using 14 day's historic data. And each algorithm will predict 1200 steps ahead. The algorithm-specific hyperparameters are fined-tuned to achieve the best accuracy and latency on the training set.} 
Table \ref{tab:perf_comp} presents the performance of each of the models on various datasets (each row is a different dataset from one region). \edit{We configured the hyperparameters as follows: window size 150, 15 epochs, batch size 768, horizon 1200, learning rate 0.001, and series length 1800.} We present the performance in terms of two metrics: Root Mean Squared Error (RMSE) and Mean Absolute Error (MAE) between the prediction and the ground truth. 
From Table \ref{tab:perf_comp} we clearly see that mWDN \cite{wang2018multilevel} outperforms other models \edit{on average}. \edit{The mWDN demonstrates strong performance, particularly in regions with larger and stable patterns, such as West US2. The TST model, requiring a longer period of input data due to their increased parameters, performs well, however, its latency is the longest among all (see Figure~\ref{fig:scaling}). In contrast, the IncepT model with a 1D-convolution layer may not be sufficiently powerful to capture the diverse patterns.}


\subsection{End-to-end Pipeline} \label{sec:expppl}
In this section, we evaluate the Pareto curve for the 2-step approach (predict future demand, and then apply SAA optimizer to the prediction) and the E2E pipeline (apply SAA optimizer to historic demand for historic optimal pool size, and then use ML to forecast). 


\begin{figure}[t!]
	\centering
		\subfloat[2-step approach]
		{\includegraphics[width=0.45\columnwidth]{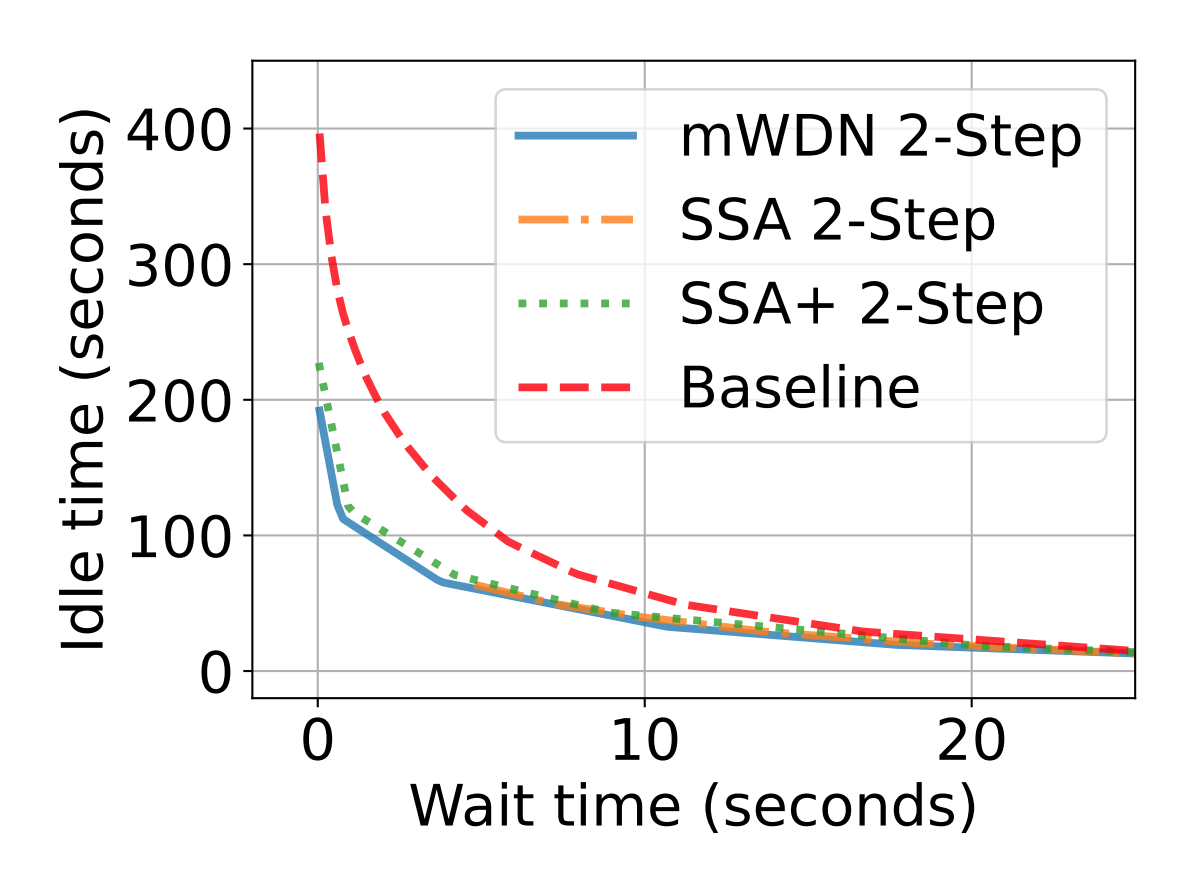}\label{fig:2step}\vspace{-0.3cm}}\hfill
		\subfloat[E2E approach]
		{\includegraphics[width=0.45\columnwidth]{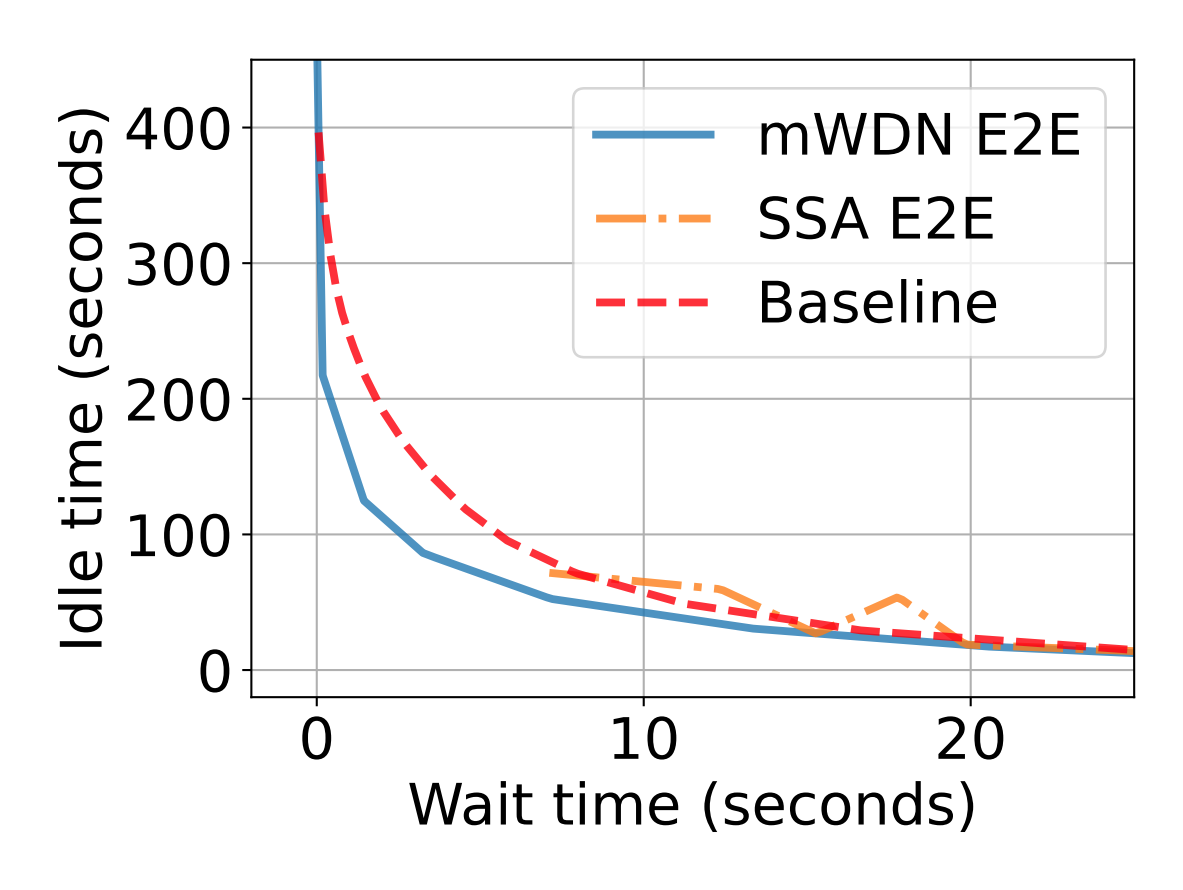}\label{fig:e2e}\vspace{-0.3cm}}
		\vspace{-0.3cm}
	\caption{Wait time vs idle time for mWDN, SSA and baseline models trained with production data.
 }
	\label{fig:trade_off_small_uswest2_idle_vs_wait}
			\vspace{-0.4cm}
\end{figure}

Figure \ref{fig:trade_off_small_uswest2_idle_vs_wait} shows the trade-off between idle time and wait time for various ML models using both the 2-step and E2E approach. ``SSA+'' denotes the results for the hybrid model. We use a no-intelligence model as baseline. A no-intelligence model's output is defined as: 
\begin{equation}
    \hat{y} = \gamma \cdot \max(y_{\text{train}})
\end{equation}
where $max(y_{\text{train}})$ is peak request rate of the training data, $\gamma$ is a fixed constant and $\hat{y}$ is the predicted cluster/session request rate. 

Figure \ref{fig:trade_off_small_uswest2_idle_vs_wait} reveals some interesting results: (1) The difference in COGS (idle time) of baseline and other ML models increases as we target lower wait times. However, this is up to a certain point, after which the difference in COGS reduces as the wait time approaches zero; (2) SSA-based models fail to achieve very low wait times (e.g., <5) for both the 2-step (in Figure~\ref{fig:2step}) and the E2E approach (Figure~\ref{fig:e2e}). However, for the mWDN model, by tuning the custom loss function (Equation \ref{eqn:loss}), we \edit{can} further increase the penalty for long wait times; (3) While we observe
that an end-to-end pipeline has better prediction performance to predict optimal pool size directly, the trade-off curve suggests 2-step performs better (see Figure~\ref{fig:2step} compared to Figure~\ref{fig:e2e}).

Targeting 99\% pool hit rate (the percentage of cluster requests experiencing 0 wait time), the system achieves up to 43\% reduction in idle time compared to static pooling. The COGS savings given different SLAs for customer wait time are shown in Table~\ref{tab:perf_comp2}. With \sysname, compared to the simple heuristics of static pooling, we are able to achieve large monetary savings.
\begin{table}[t]
\caption{Estimated annual cost savings with intelligent pooling for US (7 regions).} 
\vspace{-0.2cm}
\begin{adjustbox}{width=0.9\columnwidth}
\begin{tabular}{|c|c|c|c|c|}
\hline
Target Wait (Hit rate) & Static Pool & SSA+      & Savings SSA+ & Savings mWDN\\ \hline
0.5s ($\sim$99.9\%)      & \$>20M   & \$>15M & \$>5M &\$>5M      \\ 
1s ($\sim$99\%)          & \$>15M   & \$>10M & \$>5M &\$>5M      \\ 
5s ($\sim$95\%)          & \$>5M    & \$>5M  & \$>2M &\$>2M      \\ \hline
\end{tabular}
\end{adjustbox}
\vspace{-0.2cm}
\label{tab:perf_comp2}
\end{table}

\subsection{Data Scaling}\label{sec:scaling}

\begin{figure}[t!]
    \includegraphics[scale=0.46]{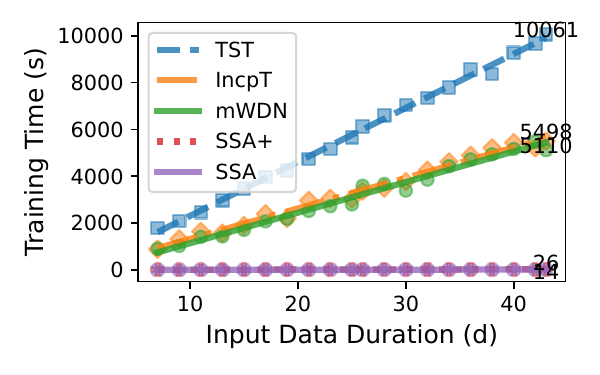}
    \vspace{-0.5cm}
    \caption{\edit{Training time vs input data size.}
    }
    \label{fig:scaling}
        \vspace{-0.5cm}
\end{figure}

We evaluated the training time of the ML models using different data sizes (see Figure~\ref{fig:scaling}). The hybrid model built on top of SSA has a slightly increased training time compared to SSA, but it is still \edit{extremely fast (200x faster) compared to the pure deep learning models (mWDN, TST or InceptionTime}). In production, we deployed the SSA+ model and trained it in an infinite loop, as it reaches similar performance as mWDN with significantly reduced latency. 
Compared with most ML pipelines where models are trained at a lower frequency and preserved into model files to be fetched at the inference time, such design significantly reduces the complexity of maintenance and development. \edit{The latency of the optimization module remains unchanged as the input data size equals the length of the predicted time frame (which is set to 1 hour for the production pipeline).}  


\subsection{Production Deployment}\label{sec:deployment}
\begin{figure}[t!]
    \includegraphics[width=0.8\columnwidth]{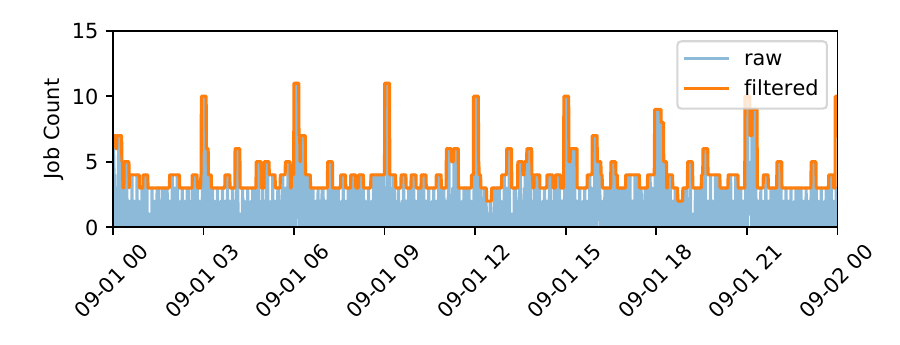}
    \vspace{-0.4cm}
    \caption{Raw versus filtered demand.}
    \label{fig:filter}
        \vspace{-0.4cm}
\end{figure}

\edit{We deployed \sysname across all production regions within \fabric in Nov 2023, with results showing great promise in significantly reducing COGS (>60\% for some production regions) and no impact on other workers co-hosted. This pipeline is scheduled to run in a continuous loop and generate pool size recommendations for the next hour.} Nevertheless, in one specific region, the ML predictions exhibit lower accuracy due to sporadic spikes occurring approximately every 3 hours (albeit not precisely timed), posing challenges for precise demand and spike arrival forecasting. To bolster ML robustness and enhance pool hit rates, we implemented the following strategies: (1) prior to ML training, we applied a max filter to smooth the time-series data based on a SMOOTHING FACTOR (SF), resulting in ``fatter'' spikes (see Figure~\ref{fig:filter}) by replacing the demand $D$ in Equation~\eqref{eq:cons1} with $\Bar{D}$ as in Equation~\eqref{eq:smooth}; (2) for the linear optimizer, we extended the $\text{STABILITY}$ period to 10 minutes, forcing \sysname to recommend advanced pool size adjustments to accommodate spikes; (3) we applied a max filter (similar to Equation~\eqref{eq:smooth}) for the recommended pool size based using $SF=\tau$ to ensure that the pool size was increased for a sufficiently long period of time for spiky demand. Utilizing all the strategies, \sysname effectively addresses demand spikes, and even if they occur irregularly, the pool size is always sufficient. These enhancements for model robustness were deployed in production, leading to a further increase in COGS savings from 18\% to 64\% by significantly reducing the pool size when demand is close to zero while maintaining the hit rate to ~100\%.


{
\footnotesize
\begin{align}
  \Bar{D}(t) =
  \begin{cases}
  \max \{ D(t-\nint{\text{SF}/2}), ...,  D(t+\nint{\text{SF}/2}))\},\quad \forall t\geq \nint{\text{SF}/2} \\
  \max \{ D(0), ...,  D(t+\nint{\text{SF}/2})\},\quad \forall t< \nint{\text{SF}/2}\label{eq:smooth}
  \end{cases}
\end{align}
}

Note that in production we:
(1) run the the pipeline in a continuous loop to update the pool size with high frequency, \edit{that requires low} the end-to-end latency of the algorithm (see Section~\ref{sec:exp});
(2) set up a guardrail to validate the ML model's prediction accuracy before running the downstream optimization;
(3) set up an alerting system for pipeline failures as well as a monitoring system such that we can be informed and investigate any potential issues. And we track the \sysname status (succeeded, failed), metrics of average idle time, recommended pool size, demand request rate, pool miss/hit count/percentage, COGS saved, hydration status such as number of clusters in provisioning/ready/targeted in real-time. This comprehensive monitoring system is an essential part of the \sysname.


\edit{
\subsection{Fault Tolerance}\label{sec:fault}
Potential system failures arise from two main sources: (1) inference pipeline failure and (2) other pooling worker issues. In terms of algorithm failure, we ensure fault tolerance by generating recommendations for the next hour for each run, while executing the algorithm at more frequent intervals, e.g., 30 min. This safeguards against a single run failure, as the system retains the previous recommendation output, albeit slightly outdated. 
Additionally, in the case of consecutive system failures, leading to missing recommendations, the inferencing reverts to default configurable values. 
A health check is maintained that tracks a pooling worker's assignment status (locked or available). This involves periodic checks to confirm consistent assignment to healthy workers, overseen by the Arbitrator service. Each pooling task is leased to a worker and undergoes refreshment upon lease expiration with periodic health checks, ensuring regular health checks for all workers involved, with prompt replacement of unhealthy ones.
}

\section{Related Work}
\label{sec:related}

Performance modeling has been used to support proactive auto-scaling of resources to meet specific service-level agreements (SLAs) or Quality of Service (QoS) requirements~\cite{ruan2015optimal,bouabdallah2016use, biswas2014automatic,khorsand2019self,rahman2018auto,chen2019cost}.
With the advent of ML algorithms, research has been done in workload-forecasting methods, specifically time-series analysis, to facilitate resource management. 
\cite{herbst2013self} proposes a workload classifier based on statistics such as maximum, coefficient of variation, etc., and enumerates over a set of time-series forecasting algorithms, selecting the most appropriate one. 
\cite{poppe2022moneyball} claims that $77\%$ of the database usage on Azure SQL Database Serverless is predictable and leverages ML predictions to proactively pause and resume databases. 
To automate the scheduling of backups for PostgreSQL and MySQL servers, Seagull~\cite{poppe2020seagull} tests different ML models (including NimbusML~\cite{nimbusml}, GluonTS~\cite{gluonTS}, and Prophet~\cite{prophet}) to forecast user load for each specific server. The system identifies low-load windows with 99\% accuracy using a simple heuristic, and this solution has been deployed across all Azure regions. However, there has been limited research focusing on proactive resource provisioning to reduce the cluster initialization latency for Spark applications.

CloudNet~\cite{wood2011cloudnet} introduced the dynamic pooling of VMs by migrating networks, disks, and memory. However, in Fabric, VM pooling is unpractical due to security and authentication issues \edit{as they need to be transfered across pooled network}. In this work, we focus on cluster/session pooling.
\section{Conclusion}
\label{sec:conclusion}

In this work, we propose “pooling” (proactively provisioning) Spark clusters and sessions to eliminate the initial startup latencies. 
Intelligent Pooling involves predicting customer demand through an innovative hybrid machine learning model and dynamically adjusting the pool size to meet customer demand while reducing extraneous COGS using linear programming with low latency and high robustness. 
Evaluated using Fabric data, the system achieves up to a 43\% reduction in idle time compared to static pooling while maintaining a 99\% pool hit rate. 
Deployed in production, the system is running robustly for several regions with significant COGS savings by capturing the spiky demand patterns with dynamic pool sizes.
Future works involve the operation of multiple pools with different configurations (cluster size, etc.).
%


%

\balance

\bibliographystyle{ACM-Reference-Format}
\bibliography{sample-base}

\end{document}